\documentclass[aip,apl,reprint,graphicx]{revtex4-1} 
\usepackage{graphicx}
\usepackage{subfigure}

\begin{document}


\title{Imaging nanoscale spatial modulation of a relativistic electron beam with a MeV ultrafast electron microscope}
\author{Chao Lu$^{1,2}$, Tao Jiang$^{1,2}$, Shengguang Liu$^{1,2}$, Rui Wang$^{1,2}$, Lingrong Zhao$^{1,2}$, Pengfei Zhu}
\affiliation{Key Laboratory for Laser Plasmas (Ministry of Education), School of Physics and Astronomy, Shanghai Jiao Tong University, Shanghai 200240, China}
\affiliation{Collaborative Innovation Center of IFSA (CICIFSA), Shanghai Jiao Tong University, Shanghai 200240, China}

\author{Yaqi Liu$^{3}$, Jun Xu}
\affiliation{School of Physics, Peking University, Beijing 100871, China}

\author{Dapeng Yu$^{3,}$}
\affiliation{Department of Physics, Southern University of Sicence and Technology, Shenzhen 518055, China}

\author{Weishi Wan}
\affiliation{School of Physical Science and Technology, ShanghaiTech Univeristy, Shanghai 201210, China}

\author{Yimei Zhu}
\affiliation{Division of Condensed Matter Physics and Materials Science, Brookhaven National Laboratory, Upton, NY 11973, USA}
\affiliation{Department of Physics and Astronomy, Stony Brook University, Stony Brook, NY 11794, USA}

\author{Dao Xiang}\email{dxiang@sjtu.edu.cn}
\affiliation{Key Laboratory for Laser Plasmas (Ministry of Education), School of Physics and Astronomy, Shanghai Jiao Tong University, Shanghai 200240, China}
\affiliation{Collaborative Innovation Center of IFSA (CICIFSA), Shanghai Jiao Tong University, Shanghai 200240, China}
\affiliation{Tsung-Dao Lee Institute, Shanghai 200240, China}

\author{Jie Zhang}
\affiliation{Key Laboratory for Laser Plasmas (Ministry of Education), School of Physics and Astronomy, Shanghai Jiao Tong University, Shanghai 200240, China}
\affiliation{Collaborative Innovation Center of IFSA (CICIFSA), Shanghai Jiao Tong University, Shanghai 200240, China}

\date{\today}

\begin{abstract}
Accelerator-based MeV ultrafast electron microscope (MUEM) has been proposed as a promising tool to study structural dynamics at the nanometer spatial scale and picosecond temporal scale. Here we report experimental tests of a prototype MUEM where high quality images with nanoscale fine structures were recorded with a pulsed $\sim$3 MeV picosecond electron beam. The temporal and spatial resolution of the MUEM operating in single-shot mode is about 4 ps (FWHM) and 100 nm (FWHM), corresponding to a temporal-spatial resolution of $4\times10^{-19}~$s$\cdot$m, about 2 orders of magnitude higher than that achieved with state-of-the-art single-shot keV UEM. Using this instrument we offer the demonstration of visualizing the nanoscale periodic spatial modulation of an electron beam, which may be converted into longitudinal density modulation through emittance exchange to enable production of high-power coherent radiation at short wavelengths. Our results mark a great step towards single-shot nanometer-resolution MUEMs and compact intense x-ray sources that may have wide applications in many areas of science.  
\end{abstract}


\maketitle 

With the advent of free-electron lasers (FELs) in the hard x-ray wavelength \cite{LCLS, SACLA, PAL}, ultrafast x-ray scattering has become the primary tool for watching atoms in motion during structural changes \cite{LCLS5years}. Due in large part to the lack of low-loss and low-aberration lens, the structure information of a non-crystalline sample in x-ray scattering is typically obtained with diffraction imaging in reciprocal space (see, e.g. \cite{CDIreview}). Because x-ray pulse width in FELs is comparable to that of the electron beam \cite{ZRH, NP, RMP1, RMP2} and electrons also have short wavelength enabling atomic resolution, it is natural to explore the feasibility of using electrons for recording atomic motions. While ultrafast electron diffraction (UED) with both keV and MeV electrons \cite{UED1, UED2, UED3, UCLA, THU, OSAKA, SJTU, BNL, PKU, SLAC, DESY} have been widely used for studies of ultrafast dynamics, one unique way of probing matter with electrons is to use them for direct imaging in real space, since electrons can be manipulated with electromagnetic lenses, a prominent advantage over x-rays.  

Indeed, since the invention of transmission electron microscope (TEM \cite{tem}) in the 1930s, direct imaging with electrons have played an important role in many research areas (see, e.g. \cite{NM}). While it has become possible to image atoms under equilibrium condition in real space with resolution well below 0.1~nm \cite{50pm} with a conventional TEM, the fact that the beam is emitted continuously limits its applications in studying structural dynamics which requires high resolution both in time and space \cite{PT}.

\begin{figure}[b]
\includegraphics[width=0.45\textwidth]{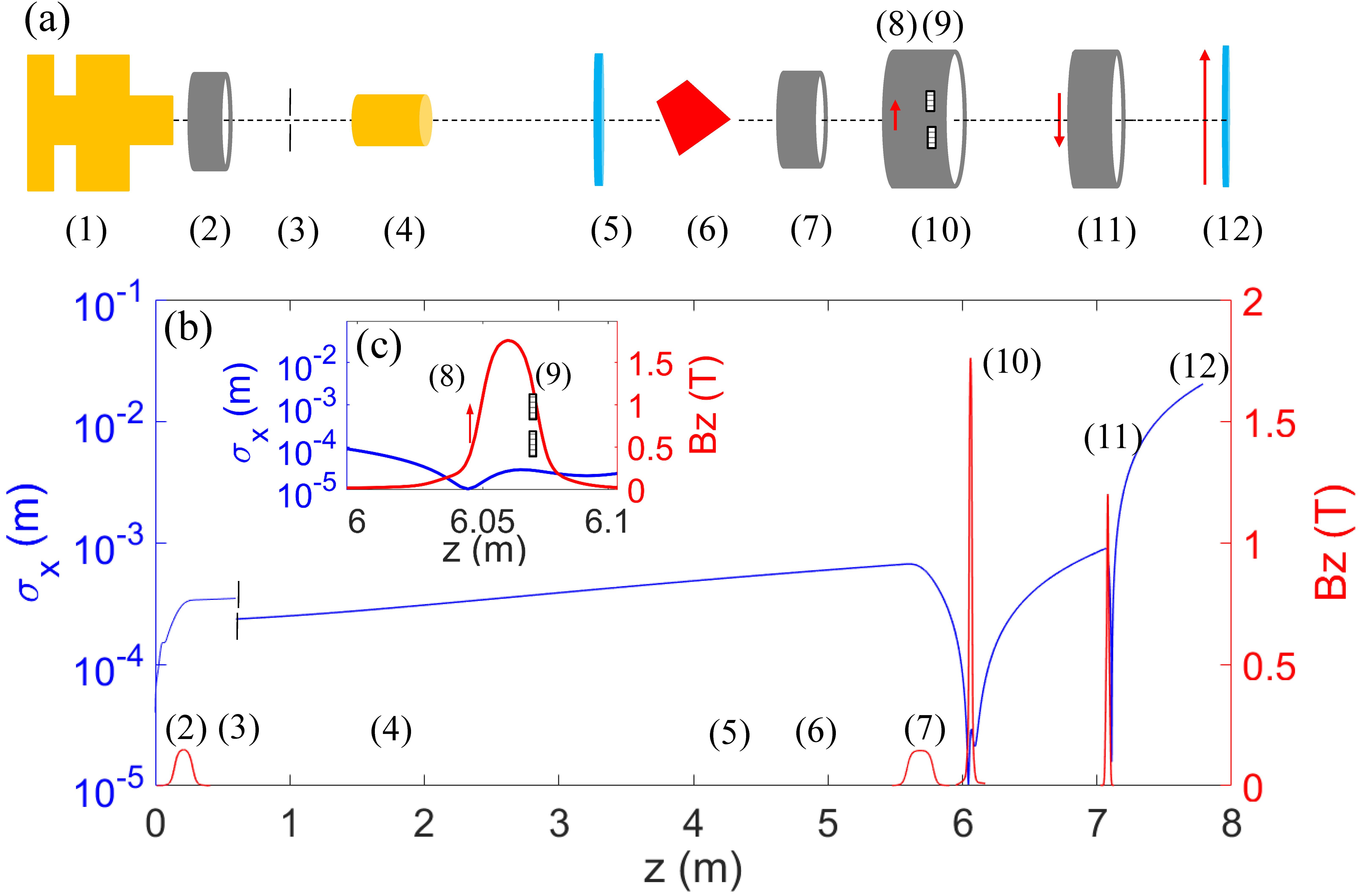}
\caption{\label{F1} (a) Schematic of the prototype MeV UEM; (b) Magnetic field distribution of the solenoid lens (red line) and the representative rms transverse beam size evolution (blue line) from the electron source to the UEM detector. The elements marked in (a) and (b) are: (1) photocathode rf gun,(2) gun solenoid lens, (3) collimator, (4) rf deflector, (5) UED detector,(6) energy spectrometer, (7) condenser lens, (8) sample, (9) objective aperture, (10) objective lens, (11) projection lens, (12) UEM detector. The inset (c) shows the beam size evolution and location of the sample and aperture in the objective lens.}
\end{figure}

The temporal resolution can be significantly improved if the electrons are bunched rather than emitted in constant stream, similar to the pulsed structure in FELs. Recently, two major configurations have been adopted to convert a TEM into an ultrafast electron microscope (UEM). The first configuration operates in single-shot mode where a useful image is obtained with a single bunch that contains millions of electrons \cite{DTEM}. With a 200~keV electron beam produced by a nanosecond laser, about 10~ns temporal resolution and 10~nm spatial resolution in a single-shot mode have been achieved \cite{DTEM1}. Analysis shows that the resolution is mainly limited by beam brightness and space charge effect. Alternatively, the UEM may operate in stroboscopic mode in which a MHz femtosecond beam with only a few electrons per pulse is used for imaging \cite{Zewail}. While this configuration indeed mitigates space charge effect and thus provides high resolution in both space and time, it is best suited for studies of reversible process.  

Increasing both the beam energy (to mitigate space charge effect) and beam brightness (to mitigate chromatic and spherical aberrations) with a photocathode radio-frequency (rf) gun has been proposed \cite{UEM1, UEM2} to break the resolution barrier in single-shot UEMs. Based on simulations, such accelerator-based MeV UEM (MUEM) may provide 10 ps temporal resolution and 10 nm spatial resolution in single-shot mode \cite{UEM1, UEM2}. In this Letter, we report on the experimental test of a single-shot prototype MUEM. Compared to imaging with a femtosecond beam (see \cite{Yang} and references therein), we used a picosecond beam that greatly mitigated the space charge effect and allowed us to image the nano-particles in a single shot. Compared to imaging with permanent magnet quadrupole lens \cite{PMQ-UCLA}, we used solenoid lenses for imaging which provide equal magnification in both horizontal and vertical directions. Furthermore, the strong-field solenoid lens with peak field approaching 2 T greatly reduced the aberration \cite{UEM1} and with a two-stage imaging system the magnification was extended to a few thousand, allowing nanoscale features to be resolved. As a representative application, we used this MUEM to visualize a free-electron crystal, i.e. electron beam with periodic spatial modulation. Through emittance exchange \cite{EEX1, EEX2, EEX3, EEX4}, the spatial modulation may be converted into longitudinal density modulation which allows the electrons to radiate in phase that may lead to orders of magnitude enhancement in radiation power \cite{ZRH, NP, RMP1, RMP2}, crucial for realizing compact intense x-ray sources \cite{Graves, Kartner, Graphene}.

The layout of the prototype MUEM is shown in Fig.~1(a). The electron beam is produced in an S-band photocathode rf gun. The instrument can be operated in either UED mode \cite{SJTU}, or in UEM mode in which the UED detector is not inserted and the solenoid lens downstream of the UED detector is used as the condenser lens which focuses the beam on the sample inside the objective lens. The beam is further magnified in the projection lens and measured with the UEM detector which consists of a phosphor screen, a lens and an Andor EM-CCD. The beam size at the sample (and thus the field of view) can be varied by tuning the strength of gun solenoid and the condenser solenoid lenses, similar to a double condenser lens system in a standard TEM. The transverse beam size evolution and locations of the key elements are shown in Fig.~1(b). 

Following the theoretical discussions \cite{UEM1, UEM2}, to mitigate space charge effect we stacked a sequence of femtosecond UV laser pulses into a nearly flat-top pulse with four $\alpha-$BBO crystals \cite{Cornell} (see Fig.~2). The laser has a Gaussian distribution in transverse direction with FWHM size of about 100 microns. In this experiment, a 0.5 mm diameter collimator is used to reduce the dark current in the gun. This also improves the quality of the image by removing the electrons with large divergence that may degrade the imaging resolution through aberrations. The beam charge before and after the collimator was measured to be about 0.40 pC and 0.20 pC, respectively. The beam charge was chosen as a compromise between beam density and beam energy spread. The electron bunch temporal profile was measured with an rf deflector (Fig.~2(d)). The beam kinetic energy is measured to be about 3.06 MeV with an energy spectrometer. The beam energy spread is measured to be about $6\times10^{-4}$ (FWHM, Fig.~2(e)) and the energy stability is about $5\times10^{-4}$ (FWHM, Fig.~2(f)). 

\begin{figure}[b]
\includegraphics[width=0.45\textwidth]{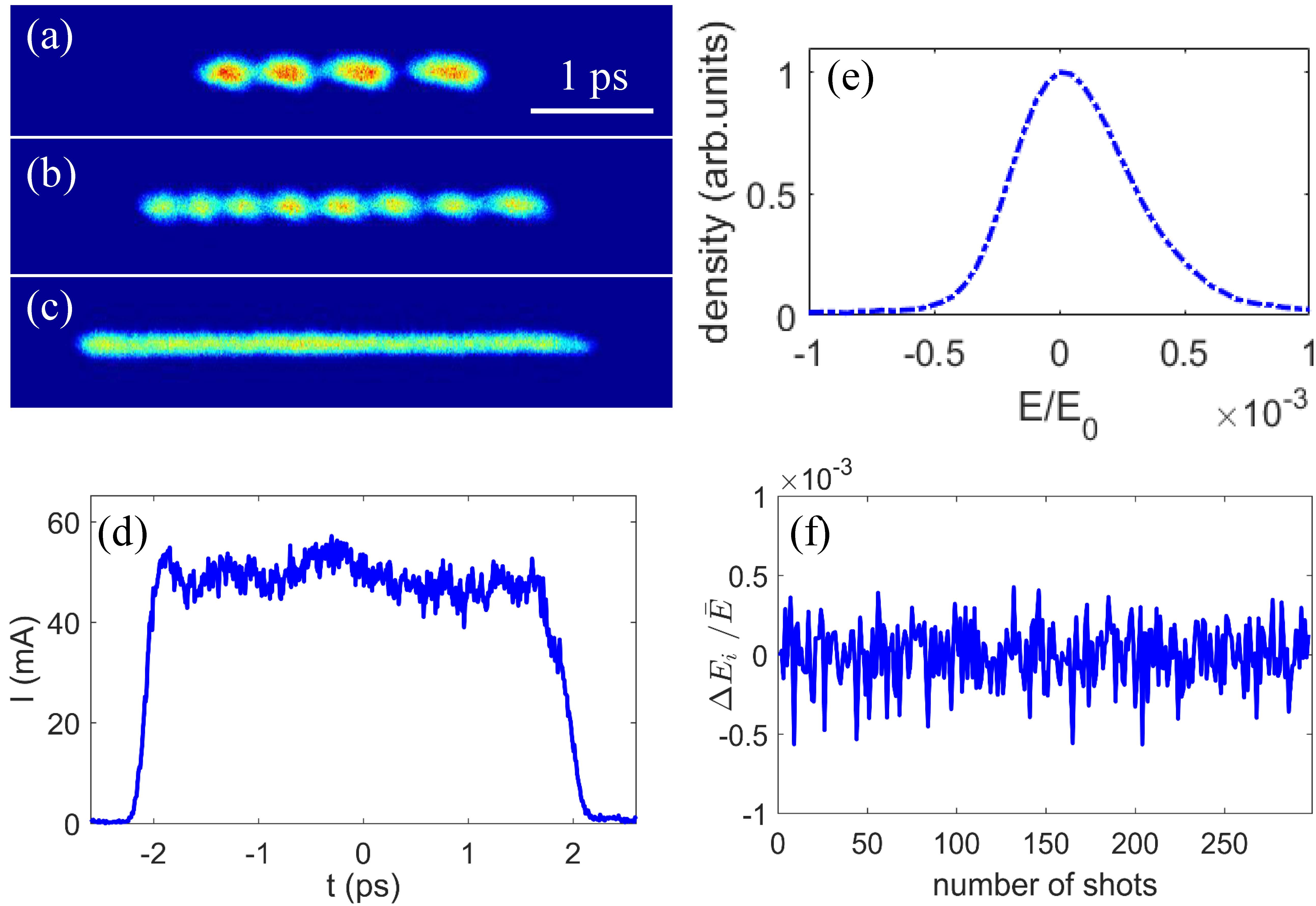}
\caption{\label{F2} Electron beam distribution measured at the UED detector with 2 $\alpha-$BBO crystals (a), 3 crystals (b), and 4 crystals (c); the corresponding beam current distribution with 4 crystals inserted is shown in (d) (bunch head to the left). (e) electron beam energy distribution; (f) electron beam energy stability.}
\end{figure}

In this experiment, a TEM copper grid (with a period of 12.5 microns and bar width of 5 microns) was first used as the sample to facilitate tuning of the MUEM. With the micron-thick grid bar, a useful image could be formed without inserting the aperture at the back focal plane of the objective lens. It is also straightforward to calibrate the magnification with the known grid dimension. After an image was formed, the objective aperture was inserted to enhance the image contrast. The representative images of the grids formed with two-stage magnification is shown in Fig.~3. In general, the contrast of the image is obtained by blocking the electrons with angles larger than that set by the objective aperture, i.e. the bright-field imaging mode (Fig~3(a)). Alternatively, we can block the electrons that pass through the sample without scattering with an annular aperture and only use those with large angles for imaging, as in the dark-field imaging mode (Fig.~3(b)). It should be pointed out that while the fine structures of the bar edge can be clearly seen in Fig.~3(a), they are less prominent in Fig.~3(b). This is likely due to the fact that in dark-field imaging, the large beam divergence degrades the image quality because of spherical and chromatic aberrations. It is worth mentioning that in this setup, the condenser lens was used to minimize the beam size at the sample to increase the beam density that generally limits the single-shot spatial resolution of UEMs \cite{UEM0, UEM1, UEM2}. With careful tuning, the minimized beam size at the sample was found to be about 8.2 microns (rms) with a beam divergence of about 1.6 mrad (rms). 

\begin{figure}[b]
\includegraphics[width=0.45\textwidth]{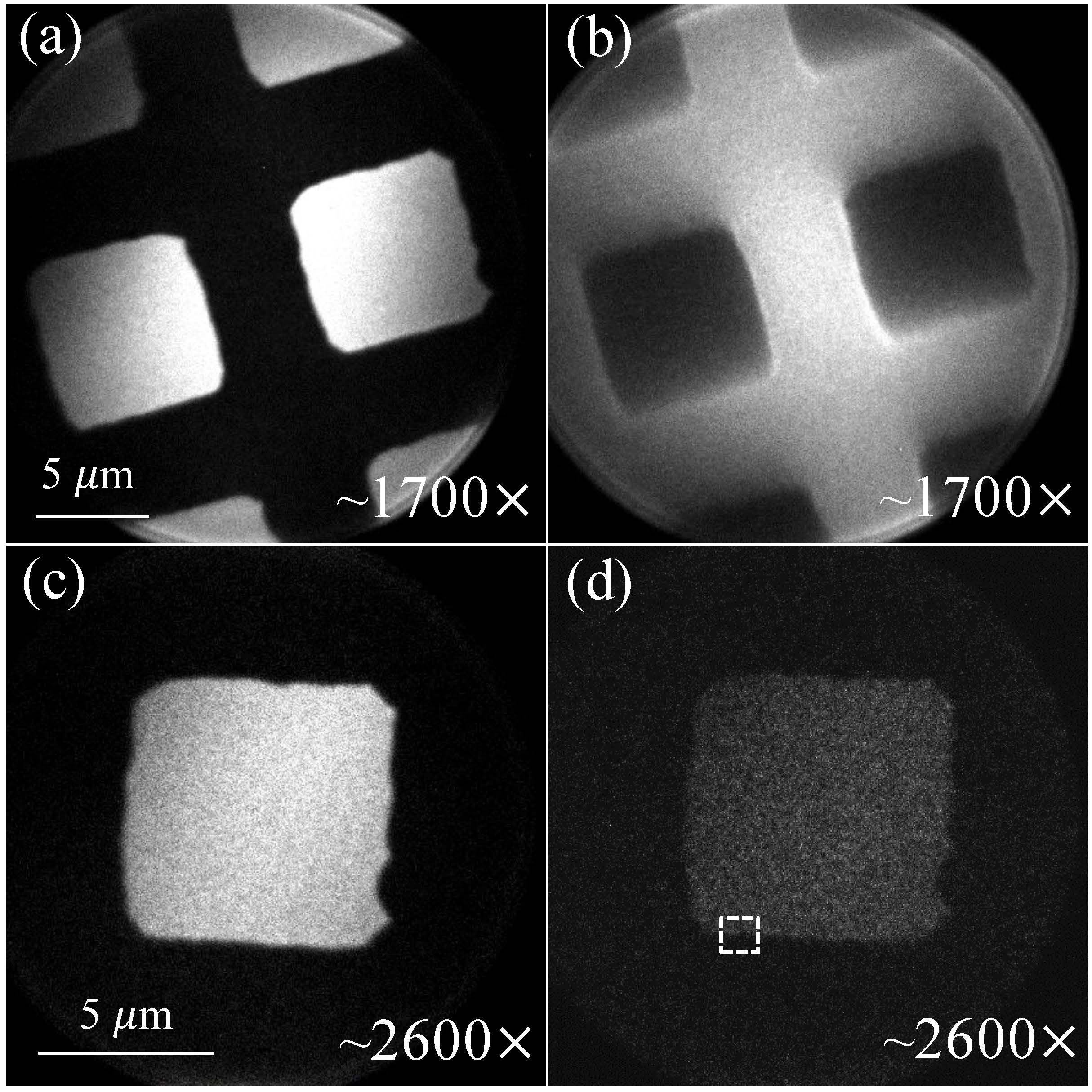}
\caption{\label{F3} Bright-field imaging (a) and dark-field imaging (b) of a 12.5 $\mu$m period TEM grid with the objective and projection lens set at 1.77 T and 1.22 T (the image was obtained with integration over 50 electron pulses); Image of the grid formed by integration over 50 electron pulses (c) and that obtained in a single shot (d) with the projection lens strength increased to about 1.50 T (the magnification is 2600).}
\end{figure}

The magnification factor of the image can be varied by tuning the strength of the projection lens. By increasing the peak field of the projection lens to 1.50 T, the magnification was increased to 2600 and the grid image formed by integration over 50 electron pulses is shown in Fig.~3(c). The curved features of the bar edge can also be seen in a single shot image (Fig.~3(d)). The projection of the grid edge (marked with white square in Fig.~3(d)) is fitted with an error function that yields a standard deviation of about 50 nm. The spatial resolution, quoted as the distance where the intensity decreases from 90\% to 10\%, is found to be about 100 nm for the single-shot mode. 

In addition to the thick sample, a holey carbon film deposited with gold nano-particles (about 400 nm diameter) was put on top of the TEM grids in a separate experiment. The image (with magnification of about 980) is shown in Fig.~4(a), where the TEM grids, the carbon film, and a gold nano-particle (marked with red circle) can all be clearly seen. The gold nano-particle was also seen in a single shot image, as shown in Fig.~4(b). To resolve finer details, the magnification factor was increased to 1900 and the image integrated over 50 pulses is shown in Fig.~4(c). With this magnification, however, the nano-particle could not be unambiguously seen in a single shot due to the low signal to noise ratio from the limited electron flux. It is worth emphasizing that the carbon foils with different thickness was also clearly differentiated in Fig.~4, e.g. the thickness for region A in Fig.~4(c) is about 3-5 nm and that for region B (the carbon support) is about 15-20 nm. This indicates that the MUEM may be readily applied to study biological samples and other low-Z materials, complementary to soft x-ray diffraction microscopy technique \cite{PNASCDI}. 

\begin{figure}[b]
\includegraphics[width=0.45\textwidth]{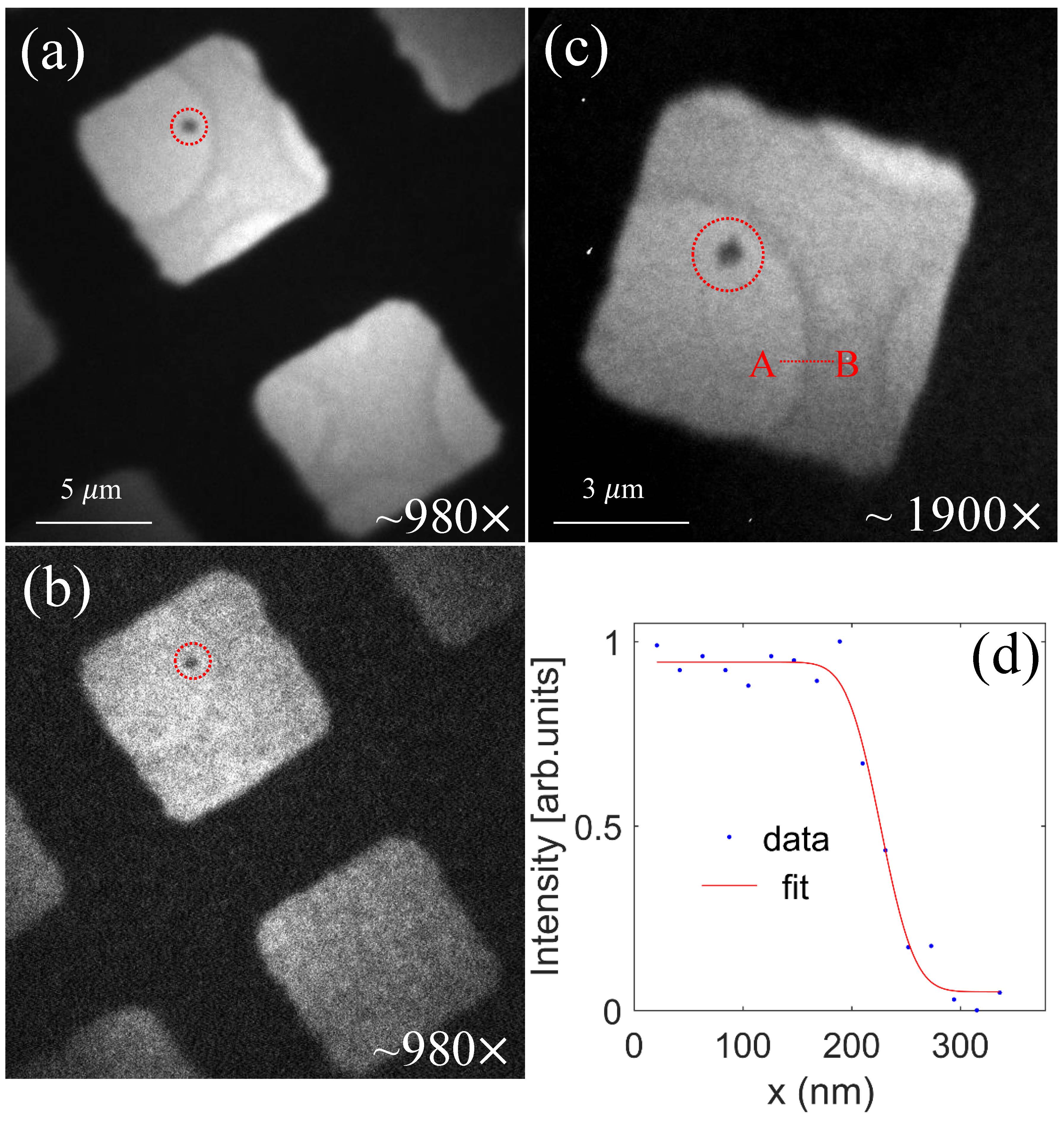}
\caption{\label{F5} Image of the carbon foil formed by integration over 50 electron pulses (a) and that obtained in a single shot (b) with magnification of 980; image of the carbon foil formed by integration over 50 electron pulses with magnification of 1900 (c) and a line-scan profile taken along the edge of the thin and thick carbon foils (d)  from which the spatial resolution of the MUEM in integration mode is estimated to be about 50 nm.}
\end{figure}

A line-scan profile (Fig.~4(d)) taken along the edge of the thin and thick carbon foils (red line in Fig.~4(c)) is used to quantify the MUEM spatial resolution in integration mode. From the fit in Fig.~4(d), the spatial resolution is found to be about 50 nm, comparable to the theoretical resolution limited by the chromatic aberration coefficient (about 1.6 cm) of the objective lens and the beam energy spread and beam divergence. The resolution for integration mode may be further improved by reducing both the beam energy spread with a harmonic cavity \cite{UEM1, UEM2} and the beam divergence with a smaller collimator. However, analysis shows that the spatial resolution for the MUEM in single shot mode in this experiment is mainly limited by the electron dose at the sample. Given the beam charge (0.2 pC) and beam size (8.2 microns rms) in this experiment, the electron density is estimated to be about 4000 electrons /$\mu$m$^2$. Assuming one needs about 40 electrons per unit sample area to form a useful image (based on Rose criterion \cite{Rose} with a contrast of 50\% and a gray scale level of 3), the single-shot spatial resolution is estimated to be about 100 nm, consistent with the experimental results.

\begin{figure}[b]
\includegraphics[width=0.45\textwidth]{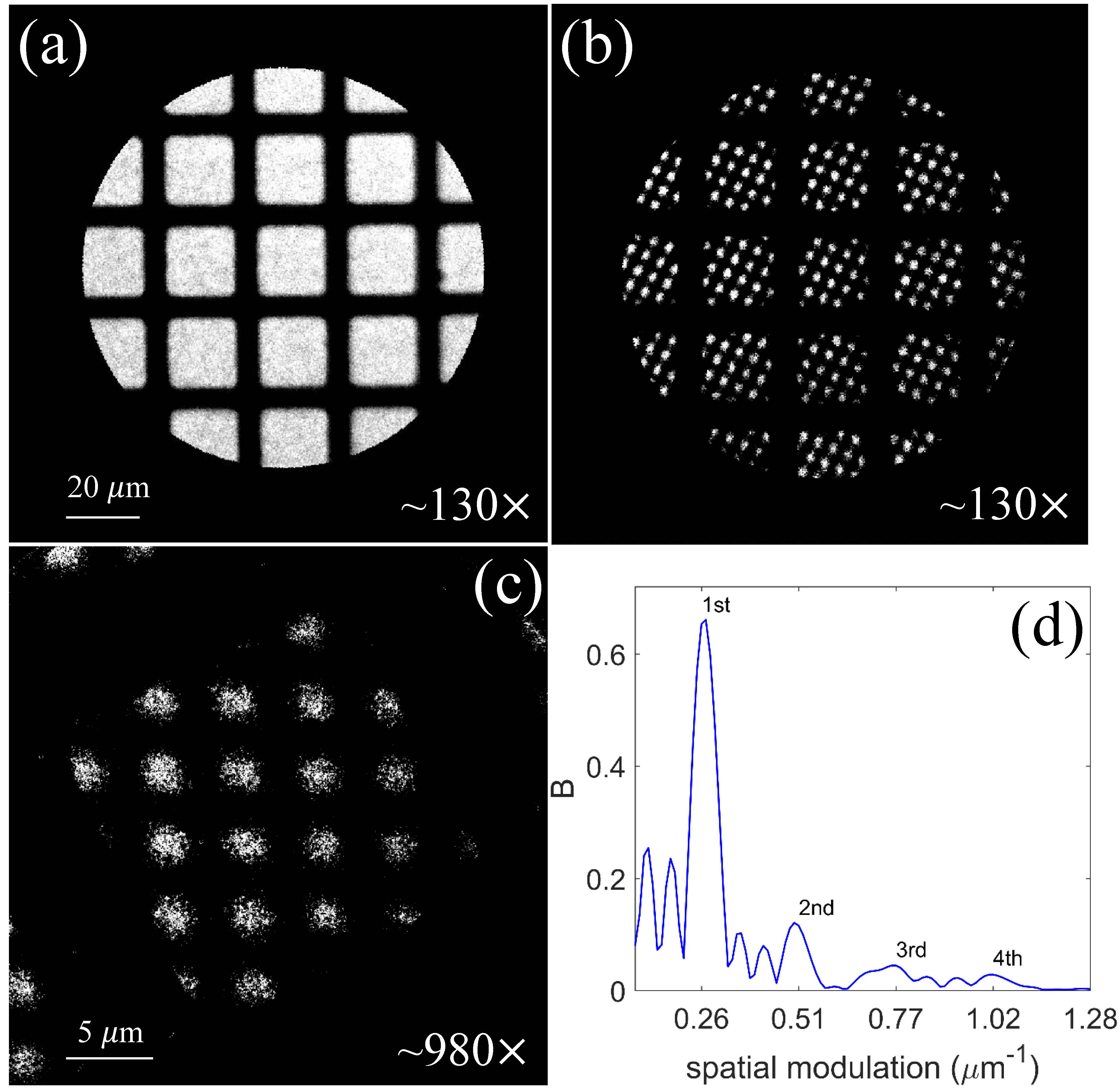}
\caption{\label{F4} Beam image when only the 25 $\mu$m period grid is inserted to the sample plane in the objective lens (a) and that obtained when a second TEM grid upstream of the condenser lens is also inserted (b). Free-electron crystal visualized with increased magnification (c) and bunching factors of the free-electron crystal at various spatial modulation frquencies (d). The harmonic number is also indicated in (d) and all the images are obtained by integration over 50 electron pulses. }
\end{figure}

While more efforts are needed to extend the single-shot resolution to the nanometer scale, the resolution achieved with this prototype UEM is sufficient for certain applications. In the following we show that this MUEM allows us to visualize a free-electron crystal, i.e. an electron beam with periodic spatial modulation. Such nanoscale spatial modulation is of great interest from the point of view of both fundamental physics and practical applications. For instance, through emittance exchange the nanoscale spatial modulation may be converted into longitudinal density modulation, which is the crucial step towards realization of intense compact x-ray sources \cite{Graves, Kartner, Graphene}. While many methods to produce such nanoscale spatial modulation have been proposed \cite{Graves, grating, Nanni}, the experimental work to visualize such structure remains largely unexplored, due in large part to the lack of instruments that are able to resolve the nanoscale structure. 

In this measurement, a 25 $\mu$m period TEM grid with a bar width of 6 microns and an aperture of 19 microns is first inserted at the sample plane for calibration of the magnification and for optimizing the imaging lens. After the grid image is formed (Fig.~5(a)), a second TEM grid with 12.5 microns period is inserted about 1 meter upstream of the condenser lens to produce spatial modulation in the beam. This modulation is relayed to the sample plane in the objective lens by the condenser lens with a demagnification of about 3. The demagnification factor is limited by the relatively large focal length of the condenser lens, and thus can be readily increased with a stronger lens. The free-electron crystal formed at the sample plane of the objective lens is visualized with the MUEM, as shown in Fig.~5(b) (low magnification) and Fig.~5(c) (high magnification with the projection lens on). The bunching factors B at various spatial modulation frequencies (k=1/$\lambda$ with $\lambda$ being the period of the modulation) calculated from the projection in Fig.~5(c) is shown in Fig.~5(d) where one can see the spatial modulation beyond 1 $\mu$m$^{-1}$ has been achieved.       

It should be pointed out that while the MUEM has $<$100 nm resolution, the condenser lens with a focal length of about 35 cm, is not able to preserve the very fine structures formed at the exit surface of the 12.5 $\mu$m period grid due to the large aberration coefficients, limiting the harmonic number to the fourth (Fig.~5(d)). This implies that it might be challenging to preserve the fine structures of a nanostructured cathode \cite{nanocathode1, nanocathode2} because of the relatively long focal length of the gun solenoid, limited by the length of the photocathode rf gun. From a practical point of view, it is beneficial to use the method demonstrated in this experiment to produce the nano-patterned beam when the beam is at relativistic energy. With this method the space charge effect that may smear out the fine structure is greatly mitigated and it also allows one to use strong lens with low aberration to significantly demagnify the modulation period to reach shorter wavelength in compact x-ray sources.

In conclusion, a single-shot 3 MeV prototype UEM with spatial resolution of about 100 nm (FWHM) and temporal resolution of about 4 ps (FWHM) was demonstrated. The temporal-spatial resolution ($4\times10^{-19}~$s$\cdot$m) is about 2 orders of magnitude higher than that achieved with state-of-the-art single-shot keV UEM \cite{DTEM1}. The achieved temporal-spatial resolution may allow one to monitor laser induced melting of nanorods and other irreversible process \cite{nanorod} in real space. With improved resolution, the MUEMs may allow one to directly probe the nucleation, growth and annihilation of the charge, spin and/or orbital ordered domains and phase transitions in quantum materials through single shot diffraction and imaging \cite{CMR}. As the first application, we used this prototype MUEM to visualize nanoscale spatial modulation of a relativistic electron beam. Our results confirm the basic concept of accelerator-based MUEM and may have a strong impact on emerging MUEMs and compact intense x-ray sources.

\begin{acknowledgments}
The authors are grateful to Q. Peng for help in design of solenoid lens. This work was supported by the Major State Basic Research Development Program of China (Grants No. 2015CB859700) and by the National Natural Science Foundation of China (Grants No. 11327902, 11504232 and 11655002). One of the authors (DX) would like to thank the support of grant from the office of Science and Technology, Shanghai Municipal Government (No. 16DZ2260200).
\end{acknowledgments}

\end{document}